\documentclass[12pt,preprint]{aastex}

\shorttitle{A Hot Helium Plasma in the GC Region}
\shortauthors{Belmont et al.}

\begin{document}
\title{A Hot Helium Plasma in the Galactic Center Region}

\author{R. Belmont and  M. Tagger}
\affil{CEA  Service d'Astrophysique, UMR \lq AstroParticules et Cosmologie\rq \\ Orme des Merisiers, 91191 Gif-sur-Yvette, France}
\email{belmont@cea.fr} \email{tagger@cea.fr}
\author{M. Muno, M. Morris, and S. Cowley}
\affil{Department of Physics and Astronomy, University of California, Los Angeles, CA 90095, USA}
\email{mmuno@astro.ucla.edu} \email{morris@astro.ucla.edu} \email{cowley@physics.ucla.edu}
\altaffiltext{1}{UMR âAstro-Particules et  Cosmologie"}

\begin{abstract}
Recent X-ray observations by the space mission Chandra confirmed the astonishing evidence for a diffuse, hot, thermal plasma at a temperature of $\sim$ 9.~$10^7$~K ($\sim $ 8~keV) found by previous surveys to extend over a few hundred parsecs in the Galactic Centre region. This plasma coexists with the usual components of the interstellar medium such as cold molecular clouds and a soft (~0.8 keV) component produced by supernova remnants, and its origin remains  uncertain. First, simple calculations using a mean sound speed for a hydrogen-dominated plasma have suggested that it should not be gravitationally bound, and thus requires a huge energy source to heat it in less than the escape time. Second, an astrophysical mechanism must be found to generate such a high temperature. No known source has been identified to fulfill both requirements. Here we address the energetics problem and show that the hot component could actually be a gravitationally confined helium plasma.  We illustrate the new prospects this opens by discussing the origin of this gas, and by suggesting possible heating mechanisms.
\end{abstract}


\section{INTRODUCTION}

Because of the strong dust absorption at optical wavelengths, X-ray astronomy is one of the most powerful tools to probe the central part of our Galaxy. For about twenty years, diffuse X-ray emission \citep*{Worall82,Warwick85} associated with intense iron emission line at 6.7 keV \citep{Koyama86} has been reported at the Galactic Centre. Although some emission is seen as far out as 4kpc, it strongly peaks in the~$\sim$~150 inner parsecs of the Galactic Centre region  \citep{Yamauchi90, Yamauchi93}. More recently, the H- and He-like K-lines of Si, S, Ar, Ca and the H-like Fe line at 6.9 keV have been resolved by ASCA \citep{Koyama96}. These lines and the associated continuum are produced by an optically thin, hot plasma, but the spectra are inconsistent with a single temperature model \citep{Koyama96}. The low-energy spectrum results mostly from a 0.8~keV plasma; but the continuum at higher energy and the 6.7 and 6.9~keV iron lines require a thermal, 8~keV plasma \citep{Muno04}. 

The temperature, patchy spatial distribution, and total energy of the soft (0.8~keV) component are compatible with a supernova origin \citep{Muno04}. Supernova remnants are the most powerful source of energy for heating the interstellar medium \citep{Schlickeiser02} and for generating plasmas in this temperature range at their shock front. The origin of the 8 keV plasma, on the other hand, is more puzzling. A simple comparison of the escape velocity at the Galactic Centre with the sound speed has suggested that it is not confined by the Galactic gravitational potential: the escape velocity estimated from different studies \citep{Sanders72, Breitschwerdt91, MWZ96} is about 1000-1200~km~s$^{-1}$ whereas,  for a plasma of solar abundances, the typical sound speed is more than 1500~km~s$^{-1}$. Assuming the gas flows out at the sound speed, the time to escape from the X-ray emitting region \citep[$\sim$70~pc,][]{Yamauchi90} is typically 4.$\times10^4$ years, orders of magnitude shorter than the radiative cooling time \citep[$\sim 10^8$~years,][]{Muno04}. The power necessary to heat this plasma over such a short time and in such a homogeneous manner exceeds by far any known or expected source and this energetics problem has led to much speculation.
For instance, the hot plasma has been suggested to be confined by a toroidal magnetic field \citep{Tanuma99}. However, besides the intrinsic difficulties of such a confinement (with respect to e.g. the Parker instability), radio observations show many magnetic non thermal filaments, the strongest of which are mostly perpendicular to the Galactic plane \citep{Yusef04}. The non thermal filaments are often considered as tracers of a pervasive, mostly poloidal magnetic field \citep{Morris96, Chandran00}, which would not confine the hot plasma. 
Furthermore, there is no adequate explanation for the high temperature of this plasma. First, the temperature of supernova remnants has never been observed to be more than~$\sim$~3~keV after, at most, a few hundred years \citep[see][for an example]{Hwang02}. The earlier phases are hotter but too short to explain the energy and smoothness of the diffuse emission at 8 keV. Second, although the hot spectral component resembles the emission from faint point sources, recent Chandra observations put stringent upper limits on the luminosity of these sources, which would thus have to be so numerous that their nature would also be a mystery. The most favoured candidates are Cataclysmic Variables (CVs), which are 10 times too rare to account for the observed emission \citep{Muno04}. Also, Chandra observations provide no evidence that the hard X-ray spectrum results from the interaction between a single-temperature soft plasma with kT=0.3 keV and nonthermal electrons \citep{Masai02}; the spectra are found to be more consistent with a thermal origin. Finally, it has been suggested that a recent strong activity of Sgr A$^*$ might have heated the plasma to a high temperature in two ways \citep{Koyama96}. But neither matter heated in the very centre 50,000 years ago and expanding across the vertical magnetic field nor a single shock starting 300 years age would have travelled out to 150~pc. Moreover, in both cases such recent heating would have generated a plasma far from ionisation equilibrium and with different ion- and electron temperatures which are not supported by the most recent observations with Chandra \citep{Muno04}

\section{A HOT HELIUM PLASMA}

Here, we propose an alternative explanation for the 8 keV component. Previous models had implicitly assumed a global behavior for a plasma of solar abundance and had compared a global sound speed with the escape speed. However, we show that this is not consistent since the medium is collisionless on the relevant timescale. 

\subsection{Ambipolar Separation of Elements}

The inferred plasma is composed of several species (electrons, protons, helium ions and other heavy ions) whose respective behaviours are different. For instance, electrons separate from ions because of their very low mass, and the resulting electrostatic field couples all the species in the plasma. The magnetic field could in principle also modify the particle dynamics, but we are interested in their vertical motion, i.e., in the geometry we consider, the direction parallel to the field, along which the field has no influence. The general case for a multi-species plasma is complex and will be discussed below, but the solution for a two-species plasma (electrons + one ion species) is enlightening. In this case, the electric field keeps ions and electrons closely coupled, transmitting the electron pressure gradient to the ions and helping them to escape. Both ions and electrons must be included in the discussion whether the plasma is bound or escaping; this can be taken into account by using a mean \lq molecular\rq~weight  $\mu = (n_i m_i + n_e m_e)/(n_i m_p+n_e m_p) $ when estimating the thermal velocity. Such a plasma is bound if the mean thermal velocity satisfies: 
$$ \bar{v}_{th}=\sqrt{\frac{k_b T}{\mu m_p}} <  v_{esc} $$
For instance, the mean thermal velocity of a pure hydrogen plasma ($\mu$=.5) at 8~keV is 1250~km~s$^{-1}$. This is larger than the escape velocity and this hydrogen plasma should escape. We thus reach the same conclusion as previous studies which assumed a hydrogen dominated plasma. On the contrary, the effective thermal velocity of a pure helium plasma ($\mu$=4/3) is 750~km~s$^{-1}$, lower than the escape speed. Such a pure helium plasma would be bound by the Galactic potential, as would pure plasmas of heavier species. 

However the hot plasma in the Galactic Center Region is composed of more than one ion species. 
If we consider for simplicity only hydrogen and helium (the case of a plasma with more species is not presented here but is straightforward and leads to the same conclusions), their respective behavior is determined by their mutual interaction. On the one hand, for a plasma of solar abundance, the few helium ions cannot influence the protons' kinetics and hydrogen ions have to escape, whatever the collisional regime is. On the other hand, the escape of helium depends on the collisions with protons: in a fully collisional regime, escaping protons exert a strong drag on helium ions which must then also escape; but in a collisionless regime, helium ions do not feel protons and remain bound in the Galactic plane. 

At the inferred number density of n$_{\rm H}$= 0.1~cm$^{-3}$, the collision time of protons escaping at their thermal velocity on helium ions for Coulomb collisions is $\tau_{\rm He/H}$= 1.$\times10^5$~years. This is substantially longer than their free escape time ($\tau_{\rm esc} \approx 4.\times10^4$~yr).  Turbulence might  cause a diffusion of the protons, and thus significantly lengthen their escape time to the point that it would become longer than their collision time with helium. However we believe that neither large-scale distortions of the magnetic field lines, nor small-scale turbulence, can be sufficiently strong to achieve this. For milligauss fields \citep{Morris96} or even fields in equipartition with the gas pressure \citep{LaRosa05}, a large scale turbulence would be very difficult to reconcile with the straightness and the vertical direction of the observed non thermal filaments. Furthermore with a vertical field, any turbulent perturbation within the disc would travel vertically out of the disc, so that contrary to the ordinary ISM, it would be very difficult to sustain a high level of turbulence. A pervasive source of turbulence of unknown origin would then be necessary up to more than 70~pc above the disc. The motion of protons relative to the bound plasma might provide a local source of small scale turbulence but again, we believe that it cannot reach a high amplitude. Indeed, contrary to e.g. the case of cosmic rays near shocks, the hydrogen ions escape with a streaming velocity close to their thermal velocity or slower; as helium and hydrogen have the same temperature and protons are rare, the stabilizing resonance of waves with helium atoms would act strongly against the destabilizing resonance with protons, so that it would be difficult for their slow escape velocity to generate turbulence and a diffusion capable of significantly lengthening their escape time. This is the reason why we consider Coulomb collisions due to a freely escaping plasma rather than assuming a diffusive escape with a coefficient of unknown origin.

The comparison of time scales lead us to suggest that protons leave this region without dragging other elements with them. This kind of selective evaporation is common in neutral stratified planetary atmospheres, but is slightly more complex here, because of the  ionized nature of the plasma. We thus simply propose that the hard emission associated with a 8~keV plasma originates from a helium plasma confined in the Galactic plane. 

In the process producing this plasma, hydrogen ions are present but escape on the short time scale discussed above. If the lifetime of helium is long enough, the density of hydrogen remains thus very low has no significant effect on the dynamics of the bound multi-component plasma. 
The equilibrium of this plasma  is governed by a full set of hydrostatic equilibrium equations coupled by the electrostatic potential. Assuming helium dominates, it is found that, without any strong mixing process, other ion species must stratify depending on their mass. The diffusion and stratification of ionized elements have also been studied in the gaseous medium of galaxy clusters \citep{Fabian77, Gilfanov84}, but there the strong turbulence may significantly inhibit the sedimentation \citep{Chuzhoy03, Chuzhoy04}.

\subsection{Revised Properties}

Because of the high temperature, hydrogen and helium are fully ionized, so no line can be observed. However, the assumption of a helium plasma changes the usual interpretation of X-ray spectra from the Galactic Center Region. Reconsidering the spectral fits of Chandra data on the diffuse emission \citep{Muno04} with this hypothesis leads to a number density for helium of n$_{\rm He}$= 0.04~cm$^{-3}$, slightly lower than the ion density (0.1~cm$^{-3}$) found in the hydrogen model. These fits also give an abundance ratio of iron relative to helium of only a third of the solar value. This may result from the source of this gas: for instance, since refractory elements in molecular clouds are mostly trapped in grains, a process which extracts the gas from these clouds before it gets heated (thus without extracting the grains) might explain a low abundance of these elements. This could be checked by considering the abundance of argon, which is not refractory and should have a normal abundance relative to helium, using its hydrogen-like K line at 3.3~keV. This line is very weak, but originates only in the 8 keV plasma, without contamination from the other phases. It is thus possible that existing and future deep observations with XMM-Newton and ASTRO-E2 could provide the statistics sufficient to constrain the abundance ratio of argon over iron (A. Decourchelle, private communication). 

The lifetime of the bound helium plasma is limited, either by radiative losses or, given the similarity between the escape and sound velocities, by other processes such as evaporation from high latitudes. Radiative losses, for instance, only require approximately 3.$\times10^{37}$~erg/s in the central $\pi*150^2*140$~parsec$^3$, i.e. less than or comparable with the expected thermal energy released by supernovae in this region of the Galaxy \citep{Yamauchi93, Muno04}. As discussed above, supernovae are unlikely to create this hot phase, but it is now easier to seek alternative energy sources. 

\subsection{Heating mechanisms}
A possibility would be to  consider the heating associated with instabilities and magnetic reconnection, especially in the transition region between the dense molecular ring at $\sim$150 pc and the central, hot region dominated by the pervasive poloidal field \citep{Chandran01}.

We also suggest that viscous friction on molecular clouds flowing toward the Galactic Center might be important so that the high viscosity of the hot plasma could permit the gravitational energy of the clouds to be tapped as they slowly spiral in toward the Galactic Centre \citep{Bally88, Miyazaki99, Oka01}. Indeed, a hot magnetized plasma is highly viscous \citep{Braginskii65}. This is however bulk viscosity, acting on compressional motions (i.e. here on the compressible part of the wake created by a molecular cloud), rather than the shear viscosity familiar in ordinary fluids. Wakes in plasmas have been extensively studied in the context of artificial satellites and of the interaction of Io with Jupiter's magnetosphere. In ordinary fluids they contain sound waves and vortices and in magnetized plasmas, they are more complex: vortices become Alfv\'en waves, and there are two magnetosonic waves. The wake thus has several components referred to as wings. The strongest are two Alfv\'en wings associated with the propagation of Alfv\'en waves, and extending on both sides of the cloud \citep{Drell65, Neubauer80}. However, because the flow around these wings is incompressible even in the non-linear regime, the bulk viscosity cannot dissipate their energy and heat the plasma. On the other hand, the wake also contains two slow magnetosonic wings \citep{Wright90, Linker91} associated with the propagation of slow magnetosonic waves. Those are compressible and thus subject to viscous damping, so that they can possibly heat the plasma from the gravitational energy released by the inflow of the clouds. The viscosity is most efficient when the collision time is comparable with the characteristic time scale of the problem, i.e., here the cloud passing time. It is interesting to note that this is exactly the case here: the parameters are such that the collision time is of the order of the interaction time for clouds of 10~parsecs moving at $\sim$100~km~s$^{-1}$ through a magnetized plasma at the inferred density and temperature.  If this is not a coincidence it might even suggest a self-regulation process, since if viscous heating became too efficient it would heat the plasma to a point where viscosity would decrease. A detailed calculation will be presented elsewhere (Belmont et al. 2005, in preparation), but we find this  mechanism efficient enough to heat the hot plasma over its radiative lifetime.

\section{CONCLUSION}

Finding that hydrogen, although it does escape from the Galaxy, is too weakly collisional to drag other elements with it, thus opens a whole line of possibilities to explain the observation of an 8~keV plasma in the Galactic Centre region. A better determination of the argon line, and planned XMM-Newton and ASTRO-E2 observations over a range of Galactic latitudes should provide a better understanding of the abundances and the vertical stratification. Astro-E2 will also be able to determine whether the plasma is in or out of equilibrium; if it confirms the thermal origin of the diffuse emission, the model of a gravitationally bound helium plasma may help to explain what otherwise is a very serious energetic quandary. 

There would remain the problem of determining the mechanism responsible for the high temperature of this plasma.  We suggest the heating might result from the viscous friction on molecular clouds flowing toward the Galactic Center. We find that the collision rate in the plasma and velocities are such that the hot plasma is near the optimum for maximising the efficiency of viscous heating, which might indicate a self-regulation process.

\acknowledgments
The authors thank Anne Decourchelle, Thomas Chust and Thierry Foglizzo for numerous and very enriching discussions.

\end{document}